\documentclass[groupedaddress,aip,adv,reprint]{revtex4-1}
\usepackage{graphicx}
\usepackage{dcolumn}
\usepackage{bm}

\begin{document}

%\preprint{AIP/123-QED}

\title[Magnetization curves and thermodynamics of two Heisenberg edge-shared tetrahedra]{Magnetization curves and low-temperature thermodynamics of two spin-1/2 Heisenberg edge-shared tetrahedra}
\author{Jozef Stre\v{c}ka}
\email{jozef.strecka@upjs.sk}
\author{Katar\'ina Kar\v{l}ov\'a}
\affiliation{Institute of Physics, Faculty of Science, P. J. \v{S}af\'arik University, \\ 
Park Angelinum 9, 040 01 Ko\v{s}ice, Slovak Republic}

\date{\today}

\begin{abstract}
A full energy spectrum, magnetization and susceptibility of a spin-1/2 Heisenberg model on two edge-shared tetrahedra are exactly calculated by assuming two different coupling constants. It is shown that a ground state in zero field is either a singlet or a triplet state depending on a relative strength of both coupling constants. Low-temperature magnetization curves may exhibit three different sequences of intermediate plateaux at the following fractional values of the saturation magnetization: 1/3-2/3-1, 0-1/3-2/3-1 or 0-2/3-1. The inverse susceptibility displays a marked temperature dependence significantly influenced by a character of the zero-field ground state. The obtained theoretical results are confronted with recent high-field magnetization data of the mineral crystal fedotovite K$_2$Cu$_3$(SO$_4$)$_3$.  
\end{abstract}

%\pacs{Valid PACS appear here}% PACS, the Physics and Astronomy
                             % Classification Scheme.
%\keywords{Suggested keywords}%Use showkeys class option if keyword
                              %display desired
\maketitle

\section{\label{sec:intro} Introduction}
  
Low-dimensional quantum magnets are subject of considerable research interest since Haldane conjectured a fundamental topological difference between ground states of antiferromagnetic Heisenberg chains with half-odd-integer and integer spins.\cite{hal18} The Haldane conjecture about a topological ground state of the integer-valued antiferromagnetic Heisenberg spin chains has been later experimentally verified in numerous nickel- and manganese-based polymeric chains.\cite{yam00} Recently, the cluster-based Haldane state has been also predicted for the natural mineral fedotovite K$_2$Cu$_3$(SO$_4$)$_3$, which is constituted by a spin-cluster chain composed of edge-shared tetrahedra.\cite{fuj18} Beforehand the Haldane phase has been theoretically predicted for the spin-$\frac{1}{2}$ Heisenberg chain of edge-sharing tetrahedra.\cite{xia95,hon00,str14} In the present work we will comprehensively examine magnetic behavior of a spin-$\frac{1}{2}$ Heisenberg model on two edge-shared tetrahedra, which has been suggested as a strongly correlated unit cell of the spin-cluster chain realized in the fedotovite.\cite{fuj18} 
	
\section{Heisenberg edge-shared tetrahedra}
\label{sec:model}

Let us consider the isotropic spin-$\frac{1}{2}$ Heisenberg model on two edge-shared tetrahedra, which is defined in a magnetic field 
through the Hamiltonian
\begin{eqnarray}
\hat{\cal H} = J_1 \sum_{i=1}^3 {\hat{\sigma}}_{1,i} \cdot {\hat{\sigma}}_{2,i} 
+ J_2 \sum_{i=1}^2 ({\hat{\sigma}}_{1,i} + {\hat{\sigma}}_{2,i}) \cdot ({\hat{\sigma}}_{1,i+1} + {\hat{\sigma}}_{2,i+1})  
- g \mu_{\rm B} B \sum_{i=1}^3 (\hat{\sigma}_{1,i}^z + \hat{\sigma}_{2,i}^z).
\label{eq:ham}
\end{eqnarray} 
\begin{figure}
\includegraphics[width=0.49\textwidth]{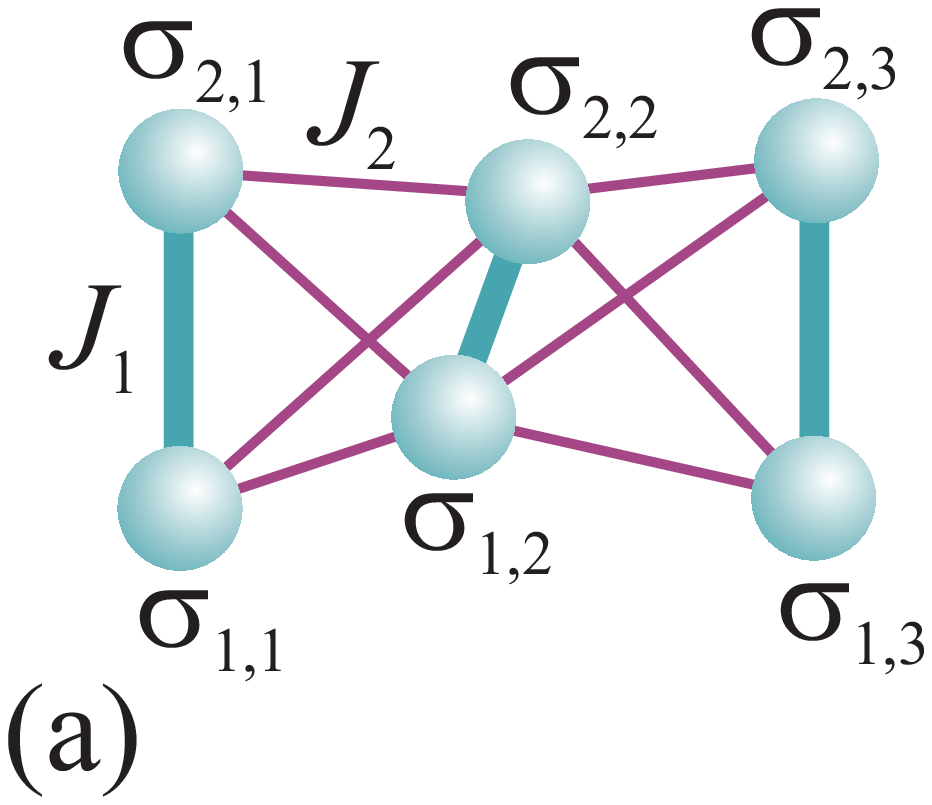} 
\hspace*{-0.5cm}
\includegraphics[width=0.5\textwidth]{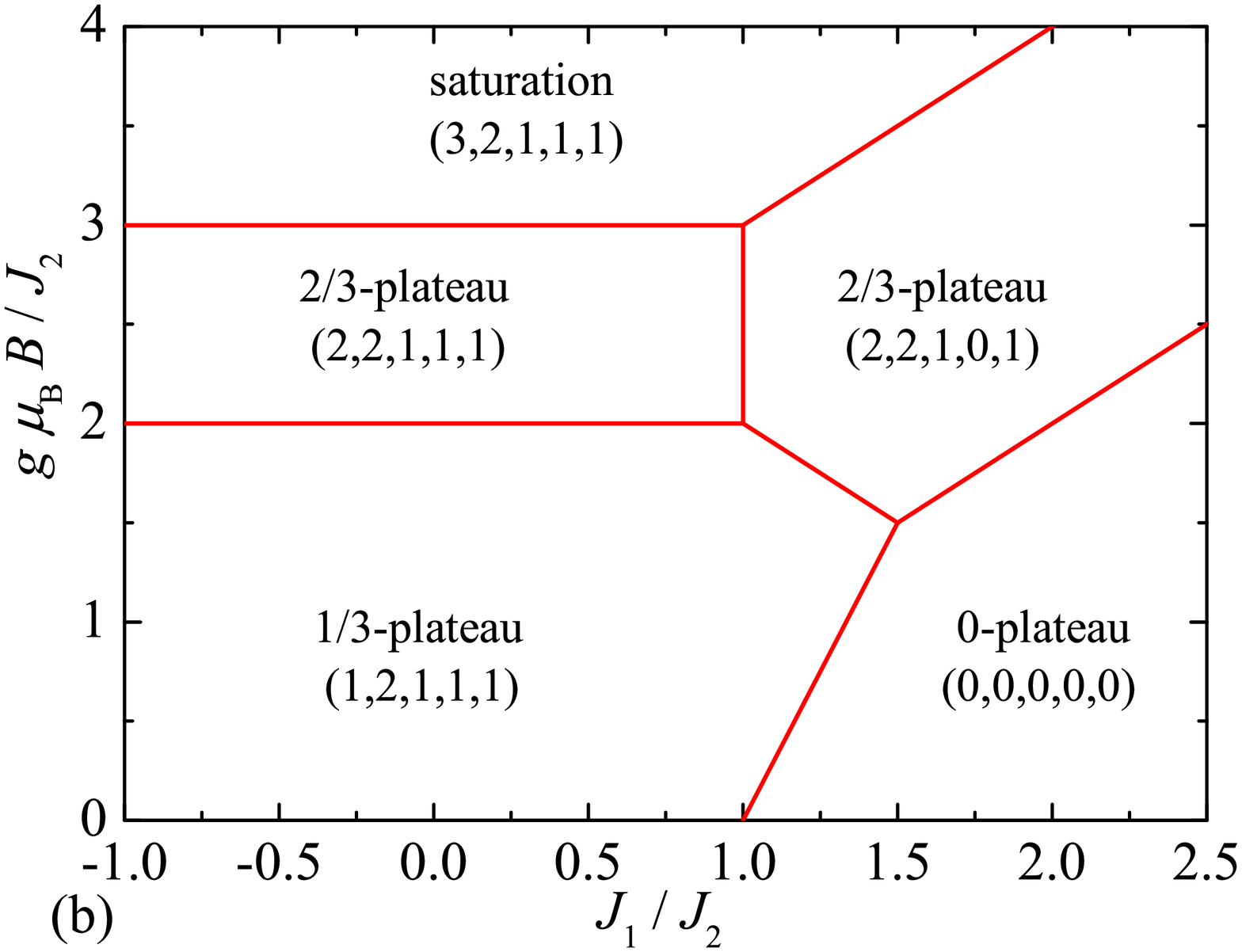} 
\vspace*{-0.5cm}
\caption{(a) A schematic illustration of two edge-shared tetrahedra. The intradimer $J_1$ and interdimer $J_2$ interactions are visualized as thick and thin lines, respectively; (b) The ground-state phase diagram in the $J_1/J_2 - g \mu_{\rm B} B/J_2$ plane. Numbers in round brackets determine particular values of the quantum spin numbers $(S_T, S_{13}, S_{1}, S_{2}, S_{3})$ within individual ground states.}
\label{fig1}
\end{figure}
The Hamiltonian (\ref{eq:ham}) involves besides the standard Zeeman's term ($g$ is gyromagnetic factor and $\mu_{\rm B}$ is Bohr magneton) three intradimer interactions $J_1$ and eight interdimer interactions $J_2$ schematically illustrated in Fig.~\ref{fig1}(a) by thick and thin lines. By introducing composite spin operators $\hat{S}_i = {\hat{\sigma}}_{1,i} + {\hat{\sigma}}_{2,i}$ for three spin pairs ($i=1,2,3$) coupled together by means of the intradimer interaction $J_1$ the Hamiltonian (\ref{eq:ham}) takes the form
\begin{eqnarray}
\hat{\cal H} = J_2 \sum_{i=1}^2 \hat{S}_{i} \cdot \hat{S}_{i+1} + \frac{J_1}{2} \sum_{i=1}^3 \hat{S}_{i}^2  
 - g \mu_{\rm B} B \sum_{i=1}^3 \hat{S}_i^z - \frac{9}{4} J_1. 
\label{eq:ha}
\end{eqnarray} 
The Hamiltonian (\ref{eq:ha}) can be subsequently put into a diagonal form upon introducing the total spin operator of the whole spin cluster $\hat{S}_T = \hat{S}_1 + \hat{S}_2 + \hat{S}_3$ and the total spin of two corner dimers $\hat{S}_{13} = \hat{S}_1 + \hat{S}_3$. Indeed, the energy eigenvalues can be then expressed in terms of quantum spin numbers $E = E(S_T,S_{13},S_1,S_2,S_3)$ of the introduced composite operators 
\begin{eqnarray}
E = \frac{J_2}{2} [S_T (S_T +1) \!-\! S_{13} (S_{13} + 1) \!-\! S_2 (S_2 +1)] 
+ \frac{J_1}{2} \!\sum_{i=1}^3 \! S_i (S_i + 1) - g \mu_{\rm B} B S_T^z - \frac{9}{4} J_1\!. 
\label{eq:es}
\end{eqnarray}
The full energy spectrum is obtained from Eq. (\ref{eq:es}) by considering all available combinations of the quantum spin numbers. The final exact result for the partition function then reads
\begin{eqnarray}
{\cal Z} &=& {\rm e}^{\frac{9}{4} K_1} + {\rm e}^{\frac{1}{4} K_1}  (1 + 2 {\rm e}^{2K_2}) + {\rm e}^{-\frac{3}{4} K_1 + 2K_2}
\nonumber \\
&+& (1 + 2 \cosh h) [3 {\rm e}^{\frac{5}{4} K_1} + {\rm e}^{\frac{1}{4} K_1}  (1 + 2 {\rm e}^{K_2}) + {\rm e}^{-\frac{3}{4} K_1} (1 + {\rm e}^{K_2} + {\rm e}^{3K_2})]
\nonumber  \\
&+& (1 + 2 \cosh h + 2 \cosh 2h) [{\rm e}^{\frac{1}{4} K_1} (1 + 2 {\rm e}^{-K_2}) + 2 {\rm e}^{-\frac{3}{4} K_1} \cosh K_2]
\nonumber \\
&+& (1 + 2 \cosh h + 2 \cosh 2h + 2 \cosh 3h) {\rm e}^{-\frac{3}{4} K_1 - 2K_2},
\label{eq:pf}
\end{eqnarray}
where $K_1 = \beta J_1$, $K_2 = \beta J_2$, $h = \beta g \mu_{\rm B} B$, $\beta = 1/(k_{\rm B} T)$, $k_{\rm B}$ is Boltzmann's constant and $T$ is the absolute temperature. The exact result for the partition function (\ref{eq:pf}) allows calculation of other important quantities such as the free energy, magnetization and susceptibility. 

\section{Results and discussion}
\label{sec:theory} 

Let us explore the most interesting results for two spin-$\frac{1}{2}$ Heisenberg edge-shared tetrahedra by assuming the antiferromagnetic interdimer interaction $J_2>0$, which has been inferred from recent magnetic measurements on the fedotovite.\cite{fuj18} At first, we will examine the ground-state phase diagram displayed in Fig.~\ref{fig1}(b) in the $J_1/J_2 - g \mu_{\rm B} B/J_2$ plane. According to this plot, the zero-field ground state is either a \textit{singlet} or a \textit{triplet} state depending on whether $J_1>J_2$ or $J_1<J_2$, respectively. While the singlet ground state causes in a zero-temperature magnetization curve 0-plateau, the triplet ground state contrarily causes $\frac{1}{3}$-plateau. Another two quintet ground states with a central dimer either in a singlet ($J_1>J_2$) or a triplet state ($J_2>J_1$) emerge at moderate magnetic fields. The quintet ground states are thereby responsible for appearance of the intermediate $\frac{2}{3}$-plateau. 

\begin{figure}
\includegraphics[width=0.49\textwidth]{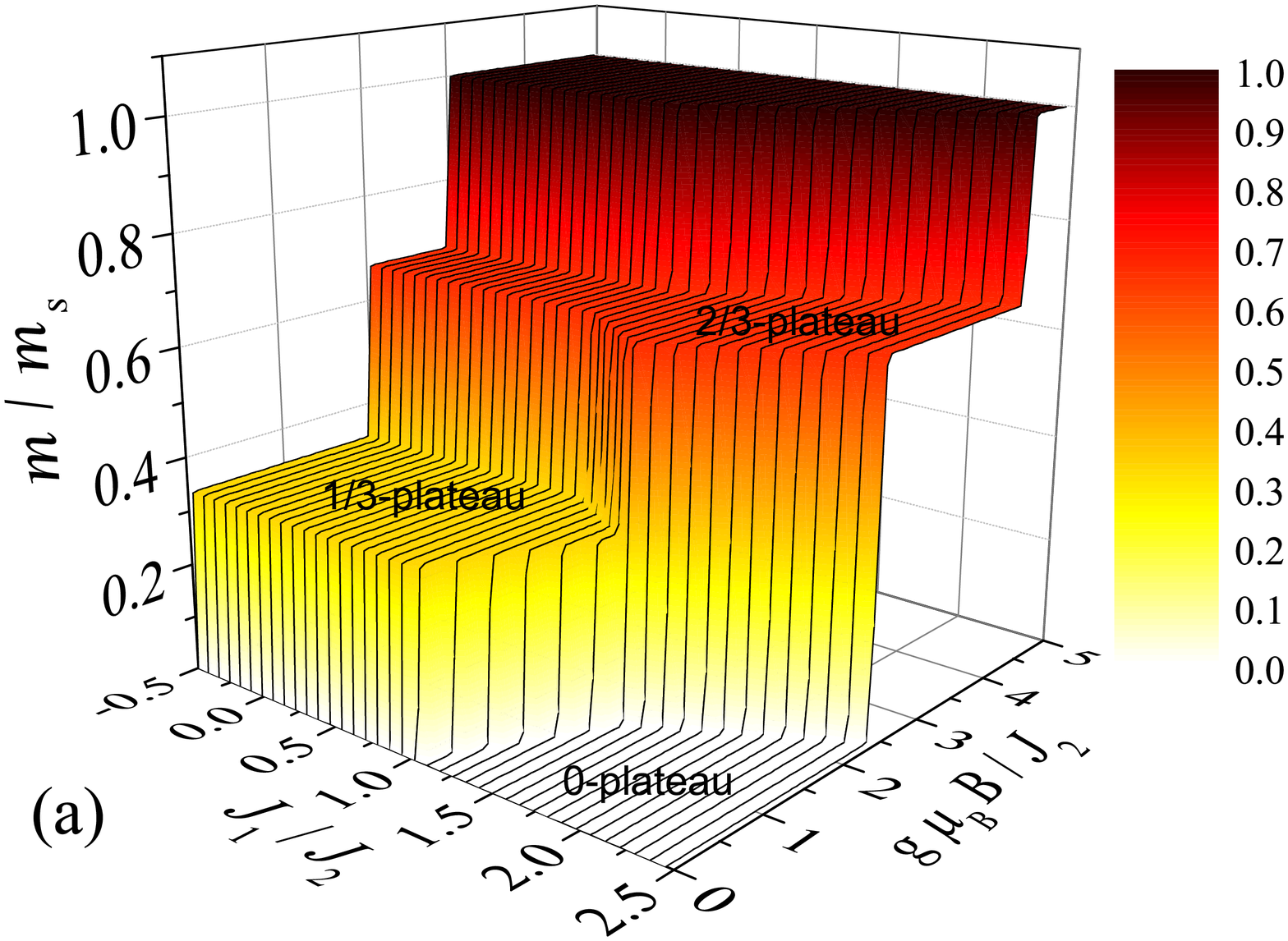} 
\includegraphics[width=0.49\textwidth]{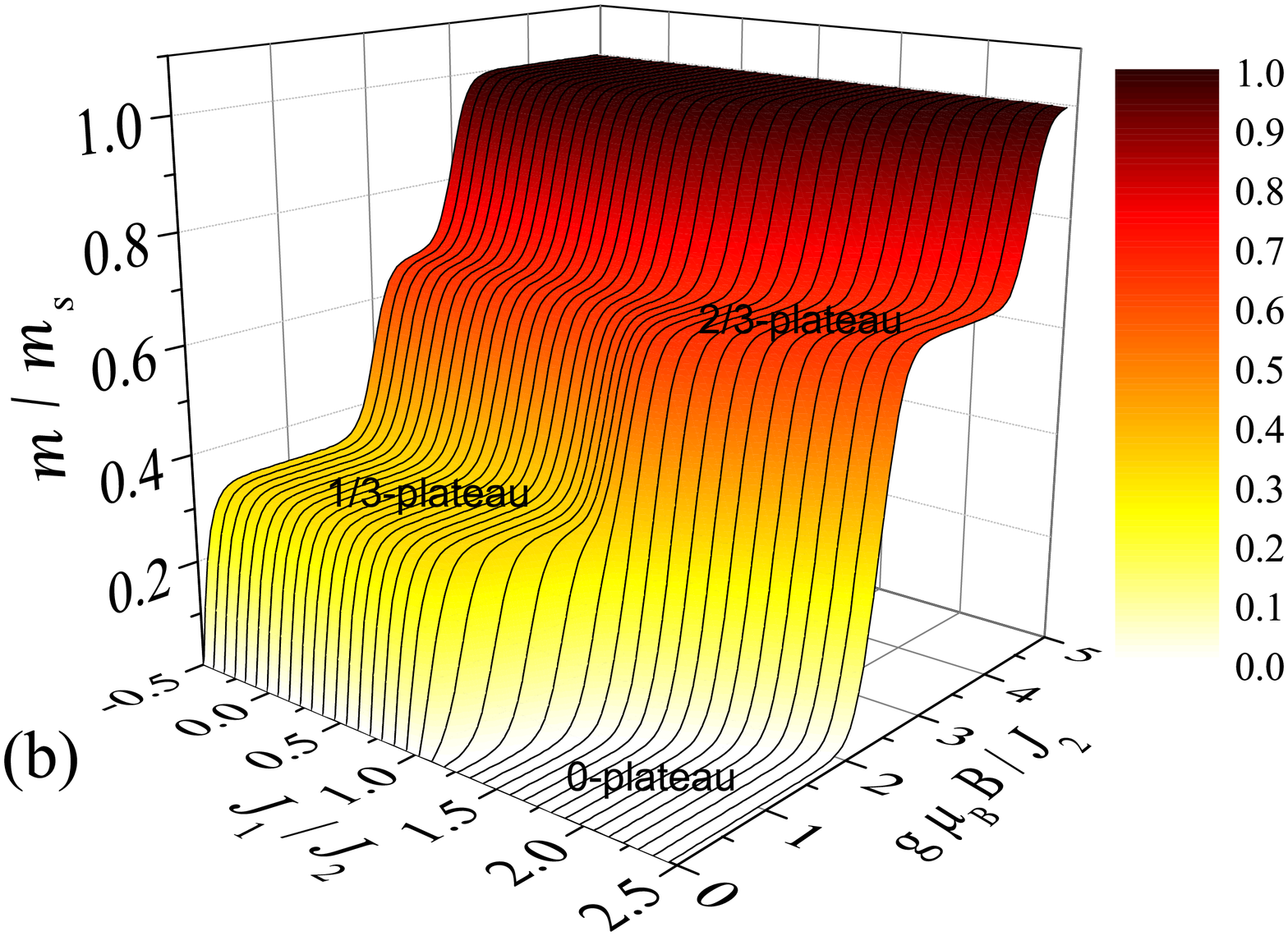} 
\vspace*{-0.5cm}
\caption{3D surface plot of the magnetization normalized with respect to its saturation value as a function of the magnetic field $g \mu_{\rm B} B/J_2$ and the interaction ratio $J_1/J_2$ at two different temperatures: (a) $k_{\rm B} T/ J_2 = 0.001$; (b) $k_{\rm B} T/ J_2 = 0.1$.}
\label{fig3}
\end{figure}

3D surface plots of the magnetization shown in Fig.~\ref{fig3} provide an independent check of the aforedescribed ground-state phase diagram. Although the magnetization does not show true magnetization jumps at any finite temperature, it exhibits at sufficiently low temperatures abrupt but continuous changes in a vicinity of each critical field being reminiscent of zero-temperature magnetization jumps. Of course, the higher the temperature is, the smoother is the respective magnetization curve due to a thermal activation of excited states. It is also evident from Fig.~\ref{fig3} that three possible sequences of the intermediate magnetization plateaux can be generally detected depending on a relative size of the interaction constants: $\frac{1}{3}$-$\frac{2}{3}$-1 for $\frac{J_1}{J_2}<1$, 0-$\frac{1}{3}$-$\frac{2}{3}$-1 for $\frac{3}{2}>\frac{J_1}{J_2}>1$ and 0-$\frac{2}{3}$-1 for $\frac{J_1}{J_2}>\frac{3}{2}$, respectively.

\begin{figure}
\includegraphics[width=0.49\textwidth]{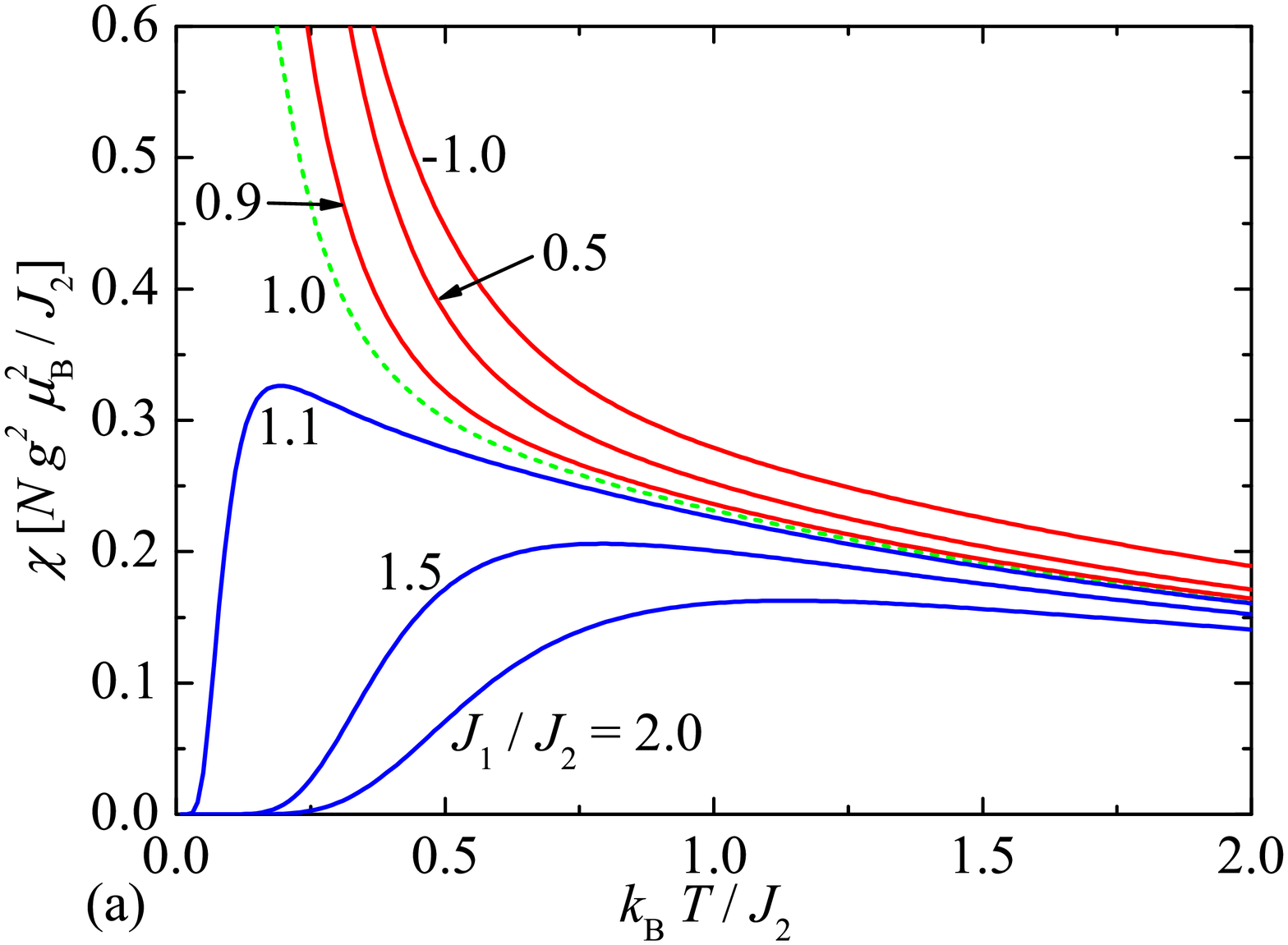} 
\includegraphics[width=0.49\textwidth]{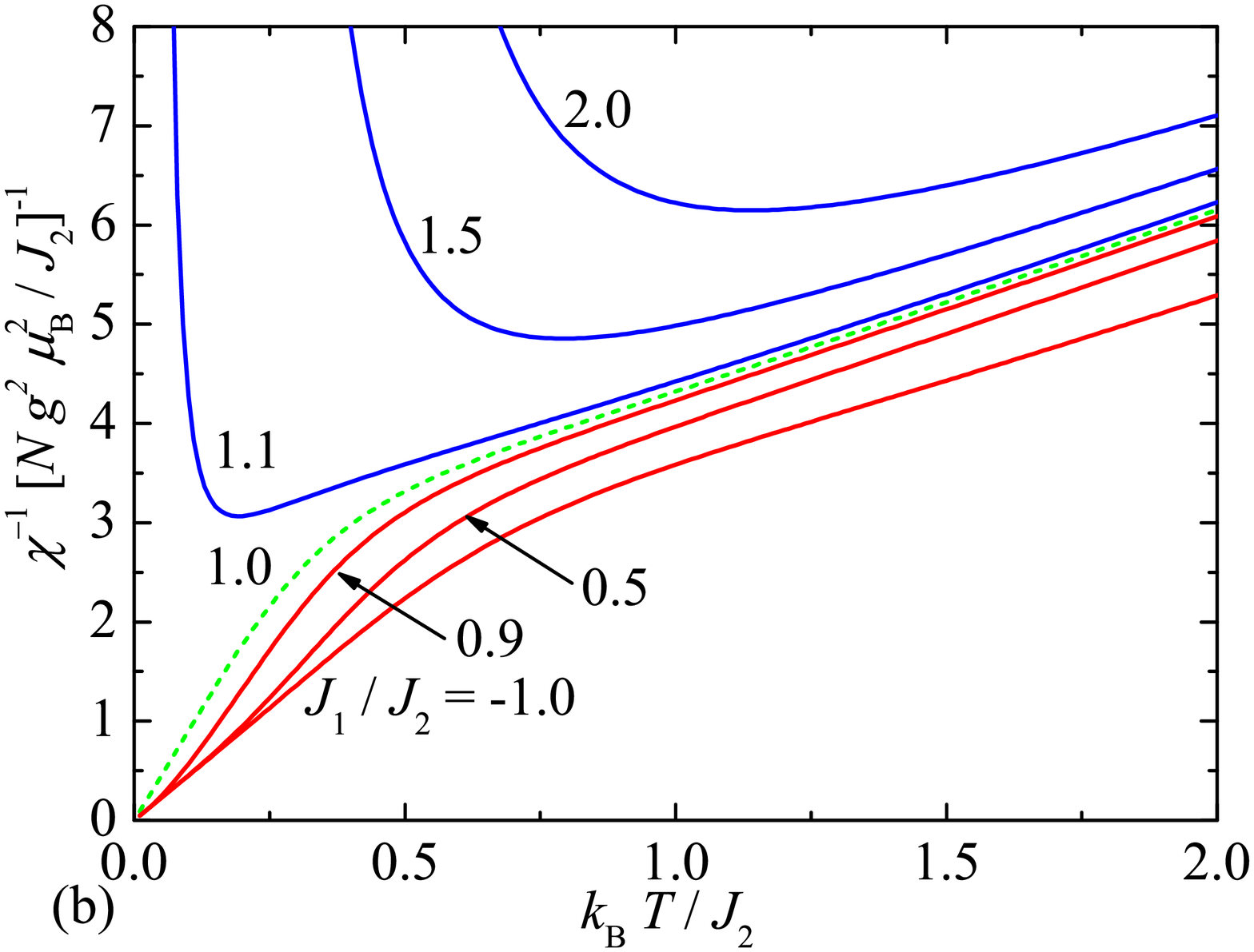} 
\vspace*{-0.6cm}
\caption{Temperature dependencies of the initial susceptibility [Fig.~\ref{fig4}(a)] and inverse initial susceptibility [Fig.~\ref{fig4}(b)] for a few selected values of the interaction ratio $J_1/J_2$.}
\label{fig4}
\end{figure}

Typical dependencies of the susceptibility are plotted in Fig.~\ref{fig4} against temperature for a few selected values of $J_1/J_2$. As could be expected, the nature of ground state basically influences a thermal behavior of the susceptibility. The susceptibility monotonically increases upon lowering temperature and it then shows a marked divergence as $T \to 0$ for $J_1/J_2<1$, whereas the inverse susceptibility has different slopes in low- and high-temperature regimes. On the contrary, the susceptibility reaches a broad local maximum upon lowering temperature and then it vanishes as $T \to 0$ for $J_1/J_2>1$, whereas the inverse susceptibility passes through a broad local minimum and then diverges as $T \to 0$. 

\begin{figure}
\includegraphics[width=0.49\textwidth]{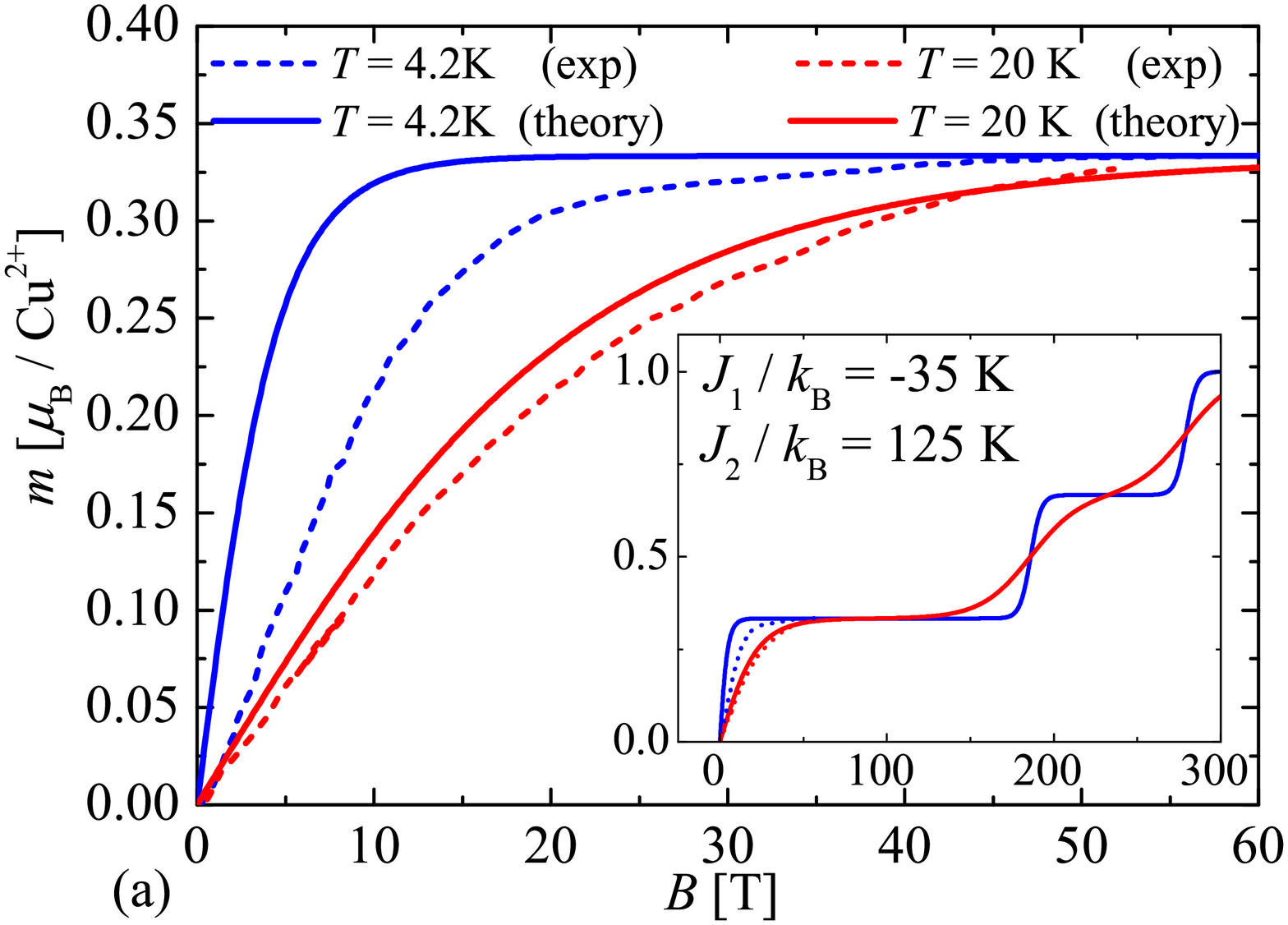} 
\includegraphics[width=0.49\textwidth]{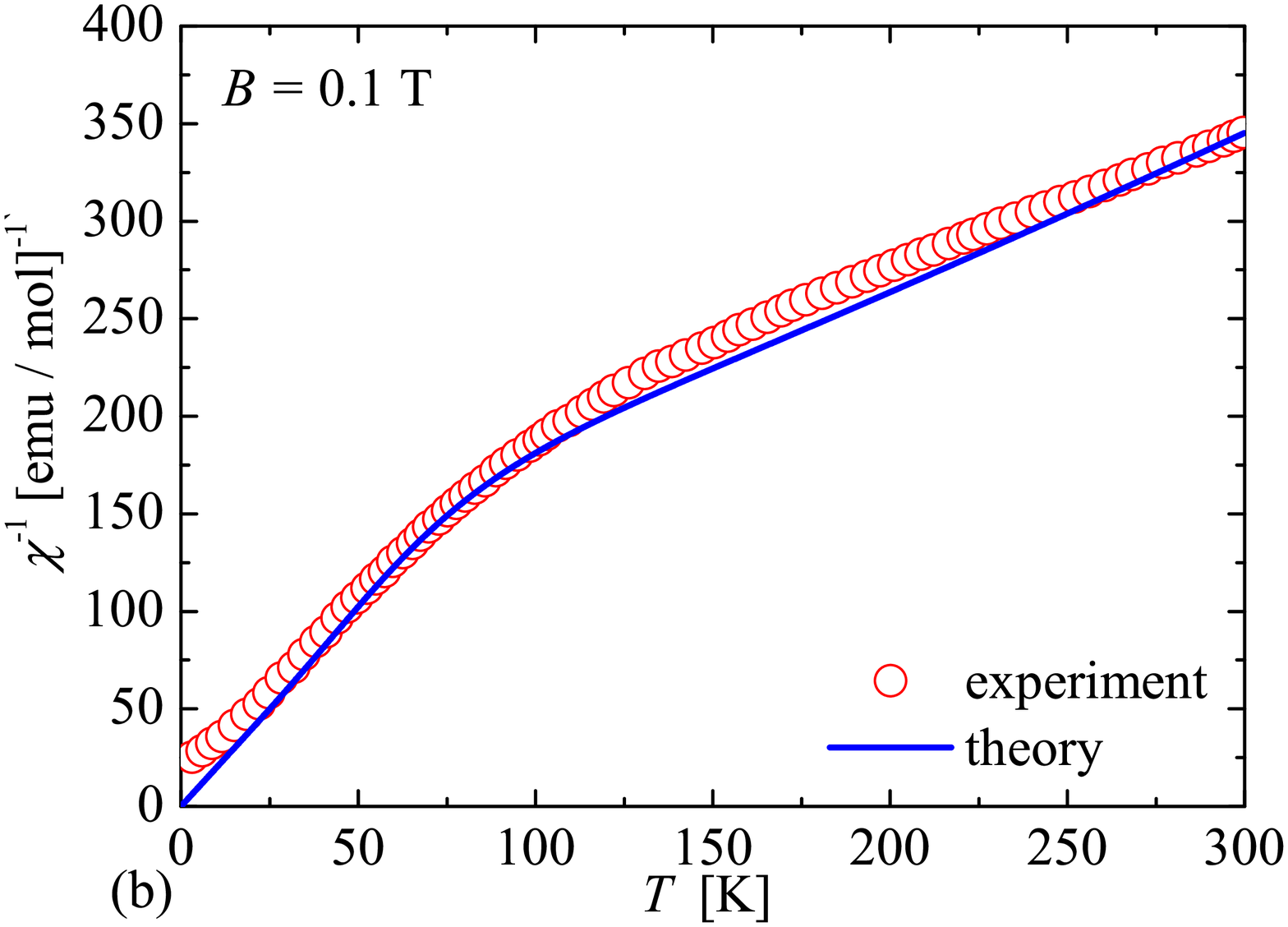} 
\vspace*{-0.7cm}
\caption{Magnetization curves and temperature dependence of the inverse susceptibility of the spin-$\frac{1}{2}$ Heisenberg model on two edge-shared tetrahedra for the coupling constants $J_1/k_{\rm B} = -35$~K, $J_2/k_{\rm B} = 125$~K and the Land\'e g-factor $g=2$ are compared with high-field magnetization (broken lines) and susceptibility data (open symbols) of K$_2$Cu$_3$(SO$_4$)$_3$  adapted according to 
Ref.~\onlinecite{fuj18}.}
\label{fig5}
\end{figure}

Finally, let us draw a comparison between theoretical results for the spin-$\frac{1}{2}$ Heisenberg model on two edge-shared tetrahedra and experimental data reported previously for the fedotovite K$_2$Cu$_3$(SO$_4$)$_3$.\cite{fuj18} High-field magnetization data of the fedotovite are compared in Fig.~\ref{fig5}(a) with the relevant theoretical results by assuming the same fitting set of the coupling constants as reported in Ref.~\onlinecite{fuj18}. Although there is a qualitative accordance between the theoretical and experimental results, the theoretical results generally precede the experimental data by reaching the $\frac{1}{3}$-plateau at lower magnetic fields. It is evident from Fig.~\ref{fig5}(a) that this quantitative discrepancy is especially marked in a magnetization curve recorded at lower temperature $T=4.2$~K. It noteworthy that any other choice of the coupling constants from the parameter space $J_2>J_1$ and $J_2/k_{\rm B} > 39$~K cannot resolve this discrepancy. The former requirement $J_2>J_1$ is inevitable in order to preserve the triplet ground state being responsible at low magnetic fields for the $\frac{1}{3}$-plateau, while the latter condition $J_2/k_{\rm B} > 39$~K shifts the critical field $g \mu_{\rm B} B_{c1} = 2 J_2$ towards the $\frac{2}{3}$-plateau (not observed in experiment) above the highest magnetic field used experimentally ($58$~T). 
%Under these constraints, the low-field part of the magnetization curve is independent of specific values of both coupling constants and the magnetization per Cu$^{2+}$ ion follows the simple formula
%\begin{eqnarray}
%m [\mu_{\rm B} / {\rm Cu}^{2+}] = \frac{g}{6} \frac{2 \sinh (\beta g \mu_{\rm B} B)}{1 + 2 \cosh (\beta g \mu_{\rm B} B)}. 
%\label{eq:me}
%\end{eqnarray}
Note that the fitting set of parameters suggested in Ref.~\onlinecite{fuj18} shifts field-driven phase transitions to the $\frac{2}{3}$- plateau and the saturation magnetization to extremely high magnetic fields $B_{c1} \approx 188$~T and $B_{c2} \approx 281$~T, respectively [see the insert in Fig.~\ref{fig5}(a)].
 
Experimental data for temperature dependence of the inverse susceptibility of the fedotovite are compared  in Fig.~\ref{fig5}(b) with the relevant theoretical prediction by assuming the same fitting set of the parameters. The theoretical curve of $\chi^{-1}$ correctly reproduces change in a slope of the experimental curve, which is almost three times greater at low temperatures than at high temperatures. The most evident disagreement between the theoretical and experimental curve can be thus found around the moderate temperature $T \approx 150$ K with the most significant change in a slope of the relevant thermal dependence.  
  
In conclusion, we have investigated a magnetic behavior of the spin-$\frac{1}{2}$ Heisenberg model on two edge-shared tetrahedra, which has been suggested as a strongly correlated unit cell of the spin-cluster chain realized in the fedotovite.\cite{fuj18} It has been evidenced that the ground state is either a singlet or a triplet state. A change in character of the ground state has substantial effect upon temperature dependence of the inverse susceptibility, which monotonically increases for the triplet ground state but shows a nonmonotonous dependence for the singlet ground state. The magnetization curves may display three possible sequences of the intermediate plateaux: $\frac{1}{3}$-$\frac{2}{3}$-1 for $\frac{J_1}{J_2}<1$, 0-$\frac{1}{3}$-$\frac{2}{3}$-1 for $\frac{3}{2}>\frac{J_1}{J_2}>1$ and 0-$\frac{2}{3}$-1 for $\frac{J_1}{J_2}>\frac{3}{2}$, respectively. Although the model brings insight into several important magnetic features of the fedotovite, it is necessarily to go beyond this simple spin-cluster model to reach a more comprehensive understanding of its complex magnetic behavior. 

\begin{acknowledgments}
This work was supported under the grants Nos. VEGA 1/0043/16 and APVV-16-0186.
\end{acknowledgments}


\begin{thebibliography}{10}
\bibitem{hal18} F.D.M. Haldane, Int. J. Mod. Phys. B 32 (2018) 1830004. 
\bibitem{yam00} M. Yamashita, T. Ishii, and H. Matsuzaka, Coord. Chem. Rev. 198 (2000) 347.
\bibitem{fuj18} M. Fujihala, T. Sugimoto, T. Tohyama and et al., Phys. Rev. Lett. 120 (2018) 077201.
\bibitem{xia95} Y. Xian, Phys. Rev. B 52 (1995) 12485.
\bibitem{hon00} A. Honecker, F. Mila, and M. Troyer, Eur. Phys. J. B 15 (2000) 227.
\bibitem{str14} J. Stre\v{c}ka, O. Rojas, T. Verkholyak, M.L. Lyra, Phys. Rev. E 89 (2014) 022143.
\end{thebibliography}
\end{document}